\documentclass[12pt]{article}
\usepackage[dvips]{graphicx}

\newcommand{\cd}{\makebox[0.08cm]{$\cdot$}}
\newcommand{\MSbar} {\hbox{$\overline{\hbox{\tiny MS}}$}}

\newcommand{\ket}[1]{\,\left|\,{#1}\right\rangle}

\begin {document}
\begin{flushright}
{\small
SLAC--PUB--10574\\
July 2004\\}
\end{flushright}

\begin{center}
{{\bf\LARGE Applications of Light-Front QCD}\footnote{Work
supported by Department of Energy contract DE--AC03--76SF00515.}}

\bigskip
{\it Stanley J. Brodsky \\
Stanford Linear Accelerator Center \\
Stanford University, Stanford, California 94309 \\
E-mail:  sjbth@slac.stanford.edu}
\medskip
\end{center}

\vfill

\begin{center}
{\bf\large Abstract }
\end{center}

Light-front Fock state wavefunctions encode the bound state
properties of hadrons in terms of their quark and gluon degrees of
freedom at the amplitude level.  The freedom to choose the
light-like quantization four-vector provides an explicitly
covariant formulation of light-front quantization and can be used
to determine the analytic structure of light-front wave functions.
The AdS/CFT correspondence of large $N_C$ supergravity theory in
higher-dimensional  anti-de Sitter space with supersymmetric QCD
in 4-dimensional space-time  has interesting implications for
hadron phenomenology in the conformal limit, including an
all-orders demonstration of counting rules for exclusive
processes.  String/gauge duality also predicts the QCD power-law
behavior of light-front Fock-state hadronic wavefunctions with
arbitrary orbital angular momentum at high momentum transfer.  The
form of these near-conformal wavefunctions can be used as an
initial ansatz for a variational treatment of the light-front QCD
Hamiltonian. I also briefly review recent work which shows that
some leading-twist phenomena such as the diffractive component of
deep inelastic scattering, single spin asymmetries, nuclear
shadowing and antishadowing cannot be computed from the LFWFs of
hadrons in isolation.

\vfill

\begin{center}
{\it Presented at  \\
QCD DOWN UNDER \\
10--13 March 2004 in the Barossa Valley \\
15--19 March 2004 at CSSM \\
Adelaide-Australia
 }\\
\end{center}

\vfill \newpage

\section{Introduction}

Light-front Fock state wavefunctions $\psi_{n/H}(x_i,\vec k_{\perp
i},\lambda_i)$ encode the bound-state quark and gluon properties of
hadrons, including their spin and flavor correlations, in the form of
universal process- and frame- independent amplitudes.  Because the
generators of certain Lorentz boosts are kinematical, knowing the
LFWFs in one frame allows one to obtain it in
any other frame.  LFWFs
underlie virtually all areas of QCD phenomenology.
The hadronic distribution amplitudes which control
hard exclusive processes are computed from the valence Fock state
LFWFs.
Matrix
elements of space-like local operators for the coupling of photons,
gravitons, and the moments of deep inelastic structure functions all can
be expressed as overlaps of light-front wavefunctions with the same number
of Fock constituents.   Similarly, the exclusive decays of heavy
hadrons such as the $B$ meson are computed from overlaps of LFWFs.  The
decays of heavy hadrons.  The unintegrated parton distributions and
generalized parton distributions measured in deeply virtual Compton
scattering can be constructed from LFWFs.  Hadronization phenomena such
as the coalescence mechanism for leading heavy hadron production are
computed from LFWF overlaps.  Diffractive jet production provides another
phenomenological window into the structure of LFWFs.  However, some
leading-twist phenomena such as the diffractive component of deep
inelastic scattering, single spin asymmetries, nuclear shadowing and
antishadowing cannot be computed from the LFWFs of hadrons in isolation.

Formally, the light-front expansion is constructed by quantizing
QCD at fixed light-cone time \cite{Dirac:1949cp} $\tau = t + z/c$
and forming the invariant light-front Hamiltonian: $ H^{QCD}_{LF} =
P^+ P^- - {\vec P_\perp}^2$
where $P^\pm = P^0 \pm P^z$~\cite{Brodsky:1997de}.  The momentum
generators $P^+$ and $\vec P_\perp$ are kinematical; $i.e.$, they are
independent of the interactions.  The generator $P^- = i {d\over d\tau}$
generates light-cone time translations, and the eigen-spectrum of the
Lorentz scalar $ H^{QCD}_{LF}$ gives the mass spectrum of the
color-singlet hadron states in QCD together with their respective
light-front wavefunctions.  For example, the proton state
satisfies: $ H^{QCD}_{LF} \ket{\psi_p} = M^2_p \ket{\psi_p}$.  The
expansion of the proton eigensolution $\ket{\psi_p}$ on the
color-singlet $B = 1$, $Q = 1$ eigenstates $\{\ket{n} \}$ of the
free Hamiltonian $ H^{QCD}_{LF}(g = 0)$ gives the light-front Fock
expansion:
\begin{eqnarray}
\left\vert \psi_p(P^+, {\vec P_\perp} )\right> &=& \sum_{n}\
\prod_{i=1}^{n} {{\rm d}x_i\, {\rm d}^2 {\vec k_{\perp i}} \over
\sqrt{x_i}\, 16\pi^3} \, 16\pi^3  \ \delta\left(1-\sum_{i=1}^{n}
x_i\right)\, \delta^{(2)}\left(\sum_{i=1}^{n} {\vec k_{\perp
i}}\right) \label{a318}\nonumber
\\
&& \rule{0pt}{4.5ex} \times \psi_{n/H}(x_i,{\vec k_{\perp i}},
\lambda_i) \left\vert n;\, x_i P^+, x_i {\vec P_\perp} + {\vec
k_{\perp i}}, \lambda_i\right>. \nonumber
\end{eqnarray}
The light-cone momentum fractions $x_i = k^+_i/P^+$ and ${\vec
k_{\perp i}}$ represent the relative momentum coordinates of the
QCD constituents.  The physical transverse momenta are ${\vec
p_{\perp i}} = x_i {\vec P_\perp} + {\vec k_{\perp i}}.$ The
$\lambda_i$ label the light-cone spin projections $S^z$ of the
quarks and gluons along the quantization direction $z$.  The
physical gluon polarization vectors $\epsilon^\mu(k,\ \lambda =
\pm 1)$ are specified in light-cone gauge by the conditions $k
\cdot \epsilon = 0,\ \eta \cdot \epsilon = \epsilon^+ = 0.$

The solutions of $ H^{QCD}_{LF} \ket{\psi_p} = M^2_p \ket{\psi_p}$
are independent of $P^+$ and ${\vec P_\perp}$;  thus given the
eigensolution Fock projections $ \langle n; x_i, {\vec k_{\perp
i}}, \lambda_i |\psi_p \rangle  = \psi_n(x_i, {\vec k_{\perp i}},
\lambda_i) ,$ the wavefunction of the proton is determined in any
frame \cite{Lepage:1980fj}.  In contrast, in equal-time quantization,  a
Lorentz boost always mixes dynamically with the interactions, so
that computing a wavefunction in a new frame requires solving a
nonperturbative problem as complicated as the Hamiltonian
eigenvalue problem itself.
The LFWFs $\psi_{n/H}(x_i,\vec k_{\perp i},\lambda_i)$
are properties of the hadron itself; they are thus universal and
process independent.

One can also define the light-front Fock expansion using a
covariant generalization of light-front time: $\tau=x \cd \omega$.
The four-vector $\omega$, with $\omega^2 = 0$, determines the
orientation of the light-front plane; the freedom to choose
$\omega$ provides an explicitly covariant formulation of
light-front quantization~\cite{cdkm}: all
observables such as matrix elements of local current operators,
form factors, and cross sections are light-front invariants --
they must be independent of $\omega_\mu.$
In recent work, Dae Sung Hwang, John Hiller, Volodya
Karmonov~\cite{Karmanov}, and I  have studied the analytic
structure of LFWFs using the explicitly Lorentz-invariant
formulation of the front form.  Eigensolutions of the
Bethe-Salpeter equation have specific angular momentum as
specified by the Pauli-Lubanski vector.  The corresponding LFWF
for an $n$-particle Fock state evaluated at equal light-front time
$\tau = \omega\cdot x$ can be obtained by integrating the
Bethe-Salpeter solutions over the corresponding relative
light-front energies.  The resulting LFWFs $\psi^I_n(x_i, k_{\perp
i})$ are functions of the light-cone momentum fractions $x_i=
{k_i\cdot \omega / p \cdot \omega}$ and the invariant mass squared
of the constituents $M_0^2= (\sum^n_{i=1} k_i^\mu)^2 =\sum_{i
=1}^n \big [\frac{k^2_\perp + m^2}{x}\big]_i$ and the light-cone
momentum fractions $x_i= {k\cdot \omega / p \cdot \omega}$ each
multiplying spin-vector and polarization tensor invariants which
can involve $\omega^\mu.$  They
are eigenstates of the Karmanov--Smirnov kinematic angular
momentum operator~\cite{ks92}.  Thus LFWFs satisfy all Lorentz
symmetries of the front form, including boost invariance, and they
are proper eigenstates of angular momentum.

\section{Light-Front Wavefunctions and QCD Phenome\-nology}

Given the light-front wavefunctions, one can compute the
unintegrated parton distributions in $x$ and $k_\perp$ which
underlie  generalized parton distributions for nonzero skewness.
As shown by Diehl, Hwang, and myself~\cite{Brodsky:2000xy},  one
can give a complete representation of virtual Compton scattering
$\gamma^* p \to \gamma p$ at large initial photon virtuality $Q^2$
and small momentum transfer squared $t$ in terms of the light-cone
wavefunctions of the target proton.  One can then verify the
identities between the skewed parton distributions $H(x,\zeta,t)$
and $E(x,\zeta,t)$ which appear in deeply virtual Compton
scattering and the corresponding integrands of the Dirac and Pauli
form factors $F_1(t)$ and $F_2(t)$ and the gravitational form
factors $A_{q}(t)$ and $B_{q}(t)$ for each quark and anti-quark
constituent.  We have illustrated the general formalism for the
case of deeply virtual Compton scattering on the quantum
fluctuations of a fermion in quantum electrodynamics at one loop.

The integrals of the unintegrated parton distributions over transverse
momentum at  zero skewness provide the helicity and transversity
distributions measurable in polarized deep inelastic experiments
\cite{Lepage:1980fj}.  For example, the polarized quark distributions at
resolution
$\Lambda$ correspond to
\begin{eqnarray}
q_{\lambda_q/\Lambda_p}(x, \Lambda) &=&  \sum_{n,q_a}
\int\prod^n_{j=1} dx_j d^2 k_{\perp j}\sum_{\lambda_i} \vert
\psi^{(\Lambda)}_{n/H}(x_i,\vec k_{\perp i},\lambda_i)\vert^2
\nonumber \\
&& \times\ \delta\left(1- \sum^n_i x_i\right) \delta^{(2)}
\left(\sum^n_i \vec k_{\perp i}\right) \delta(x - x_q)\nonumber \\
&& \times\  \delta_{\lambda_a, \lambda_q} \Theta(\Lambda^2 - {\cal
M}^2_n)\ , \nonumber
\end{eqnarray}
where the sum is over all quarks $q_a$ which match the quantum
numbers, light-cone momentum fraction $x,$ and helicity of the struck
quark.
As shown by Raufeisen and myself~\cite{Raufeisen:2004new}, one can construct
a ``light-front density matrix" from the complete set of light-front
wavefunctions which is a Lorentz scalar.  This form can be used at finite
temperature to give a boost invariant formulation of thermodynamics.  At zero
temperature the light-front density matrix is directly connected to the
Green's function for quark propagation in the hadron as well as deeply
virtual Compton scattering.  In addition, moments of transversity
distributions and off-diagonal helicity convolutions are defined from the
density matrix of the light-cone wavefunctions.
The light-front wavefunctions also specify the multi-quark and
gluon correlations of the hadron.  For example,  the distribution
of spectator particles in the final state which could be measured
in the proton fragmentation region in deep inelastic scattering at
an electron-proton collider are in principle encoded in the
light-front wavefunctions.

Given the $\psi^{(\Lambda)}_{n/H},$ one can construct any
spacelike electromagnetic, electroweak, or gravitational form
factor or local operator product matrix element of a composite or
elementary system from the diagonal overlap of the LFWFs~\cite{BD80}.
Exclusive semi-leptonic $B$-decay amplitudes involving timelike currents
such as $B\rightarrow A
\ell \bar{\nu}$ can also be evaluated exactly in the light-front
formalism~\cite{Brodsky:1998hn}.  In this case, the timelike decay
matrix elements require the computation of both the diagonal
matrix element $n \rightarrow n$ where parton number is conserved
and the off-diagonal $n+1\rightarrow n-1$ convolution such that
the current operator annihilates a $q{\bar{q'}}$ pair in the
initial $B$ wavefunction.  This term is a consequence of the fact
that the time-like decay $q^2 = (p_\ell + p_{\bar{\nu}} )^2 > 0$
requires a positive light-cone momentum fraction $q^+ > 0$.
Conversely for space-like currents, one can choose $q^+=0$, as in
the Drell-Yan-West representation of the space-like
electromagnetic form factors.
The light-front Fock representation thus provides an exact formulation
of current matrix elements of local operators.  In contrast, in
equal-time Hamiltonian theory, one must evaluate connected
time-ordered diagrams where the gauge particle or graviton couples
to particles associated with vacuum fluctuations.  Thus even if
one knows the equal-time wavefunction for the initial and final
hadron, one cannot determine the current matrix elements.  In the
case of the covariant Bethe-Salpeter formalism, the evaluation of
the matrix element of the current requires the calculation of an
infinite number of irreducible diagram contributions.
One can
also prove that the anomalous gravitomagnetic moment
$B(0)$ vanishes for any composite system~\cite{Brodsky:2000ii}.  This
property follows directly from the Lorentz boost properties of the
light-front Fock representation and holds separately for each Fock state
component.

One of the central issues in the analysis of fundamental hadron
structure is the presence of non-zero orbital angular momentum in
the bound-state wave functions.  The evidence for a ``spin crisis"
in the Ellis-Jaffe sum rule signals a significant orbital
contribution in the proton wave
function~\cite{Jaffe:1989jz,Ji:2002qa}.  The Pauli form factor of
nucleons is computed from the overlap of LFWFs differing by one
unit of orbital angular momentum $\Delta L_z= \pm 1$.  Thus the
fact that the anomalous moment of the proton is non-zero requires
nonzero orbital angular momentum in the proton
wavefunction~\cite{BD80}.  In the light-front method, orbital
angular momentum is treated explicitly; it includes the orbital
contributions induced by relativistic effects, such as the
spin-orbit effects normally associated with the conventional Dirac
spinors.

\section{Complications from Final State Interactions}

It is usually assumed---following the parton model---that the
leading-twist structure functions measured in deep  inelastic
lepton-proton scattering are simply the probability distributions for
finding quarks and gluons in the target nucleon.  In fact, gluon  exchange
between the fast, outgoing quarks and the target spectators
effects the leading-twist structure functions in a profound way,
leading to  diffractive leptoproduction processes, shadowing of
nuclear structure  functions, and target spin asymmetries.  In
particular, the final-state  interactions from gluon exchange lead
to single-spin asymmetries in  semi-inclusive deep inelastic
lepton-proton scattering which are not  power-law suppressed in
the Bjorken limit.

A new understanding of the role of final-state interactions in
deep inelastic scattering has recently
emerged~\cite{Brodsky:2002ue}.  The final-state interactions from
gluon exchange between the outgoing quark and the target spectator
system lead to single-spin asymmetries in semi-inclusive deep
inelastic lepton-proton scattering at leading twist in
perturbative QCD; {\em i.e.}, the rescattering
corrections of the struck quark with the target spectators are not
power-law suppressed at large photon virtuality $Q^2$ at fixed
$x_{bj}$~\cite{Brodsky:2002cx}  The final-state interaction from
gluon exchange occurring immediately after the interaction of the
current also produces a leading-twist diffractive component to
deep inelastic scattering $\ell p \to \ell^\prime p^\prime X$
corresponding to color-singlet exchange with the target system;
this in turn produces shadowing and anti-shadowing of the nuclear
structure functions~\cite{Brodsky:2002ue,Brodsky:1989qz}.  In
addition, one can show that the pomeron structure function derived
from diffractive DIS has the same form as the quark contribution
of the gluon structure function~\cite{BHIE}.  The final-state
interactions occur at a light-cone time $\Delta\tau \simeq 1/\nu$
after the virtual photon interacts with the struck quark,
producing a nontrivial phase.  Thus none of the above phenomena is
contained in the target light-front wave functions computed in
isolation.   In particular, the shadowing of nuclear structure
functions is due to destructive interference effects from
leading-twist diffraction of the virtual photon, physics not
included in the nuclear light-front wave functions.  Thus the
structure functions measured in deep inelastic lepton scattering
are affected by final-state rescattering, modifying their
connection to light-front probability distributions.  Some of
these results can be understood by augmenting the light-front wave
functions with a gauge link, but with a gauge potential created by
an external field created by the virtual photon $q \bar q$ pair
current~\cite{Belitsky:2002sm}.  The gauge link is also process
dependent~\cite{Collins:2002kn}, so the resulting augmented LFWFs
are not universal.

The shadowing and antishadowing of nuclear structure functions in
the Gribov-Glauber picture is due to the destructive and
constructive coherence, respectively, of amplitudes arising from
the multiple-scattering of quarks in the nucleus.  The effective
quark-nucleon scattering amplitude includes Pomeron and Odderon
contributions from multi-gluon exchange as well as Reggeon quark
exchange contributions~\cite{Brodsky:1989qz}.  The multiscattering nuclear
processes from Pomeron, Odderon and pseudoscalar Reggeon exchange
leads to shadowing and antishadowing of the electromagnetic
nuclear structure functions in agreement with measurements.  This
picture leads to substantially different nuclear effects for
charged and neutral currents, particularly in anti-neutrino
reactions, thus affecting the extraction of the weak-mixing angle
$\sin^2\theta_W$ and the constant $\rho_o$ which are determined
from the ratio of charged and neutral current deep inelastic from
neutrino and anti-neutrino scattering.  In recent work, Schmidt, Yang, and
I~\cite{BSY} find that a substantial part of the difference
between the standard model prediction and the anomalous NuTeV
result~\cite{Zeller:2001hh} for $\sin^2\theta_W$ could be due to the
different behavior of nuclear antishadowing for charged and neutral
currents.  Detailed measurements of the nuclear dependence of charged,
neutral and electromagnetic DIS processes are needed to establish
the distinctive phenomenology of shadowing and antishadowing and
to make the NuTeV results definitive.

\section{Other Aspects of Light-Front Wavefunction Phenomenology}

A number of other important phenomenological properties follow
directly from
the  structure of light-front wavefunctions in gauge theory.

(1) Color transparency.  The small transverse size fluctuations of a
hadron wavefunction with a small color dipole moment will have
minimal interactions in a nucleus~\cite{Bertsch:1981py,Brodsky:1988xz}.
This has been verified in the case of diffractive
dissociation  of a high energy pion into dijets $\pi A \to q \bar q
A^\prime$ in which the nucleus is left in its ground
state~\cite{Ashery:2002jx}.  When the hadronic jets have balancing but
high transverse momentum, one studies the small size fluctuation of the
incident pion.  The diffractive dissociation cross section is found to be
proportional to $A^2$ in agreement with the color transparency prediction.
Color transparency has also been observed in diffractive
electroproduction of
$\rho$ mesons~\cite{Borisov:2002rd} and in quasi-elastic $p A \to p p A-1$
scattering~\cite{Aclander:2004zm} where only the small size fluctuations
of the hadron wavefunction enters the hard exclusive scattering
amplitude.  In the latter case an anomaly occurs at $\sqrt s \simeq 5 $
GeV, most likely signaling a resonance effect
at the charm threshold~\cite{Brodsky:1987xw}.

(2) Intrinsic charm~\cite{Brodsky:1980pb}.  The probability for  Fock states of a light
hadron such as the proton to have an extra heavy quark pair decreases as $1/m^2_Q$ in
non-Abelian gauge theory~\cite{Franz:2000ee,Brodsky:1984nx}.  The relevant matrix element
is the cube of the QCD field strength $G^3_{\mu \nu}.$ This is in contrast to abelian
gauge theory where the relevant operator is $F^4_{\mu \nu}$ and the probability of
intrinsic heavy leptons in QED bound state is suppressed as $1/m^4_\ell.$  The intrinsic
Fock state probability is maximized at minimal off-shellness.  The maximum probability
occurs at $x_i = { m^i_\perp /\sum^n_{j = 1} m^j_\perp}$; i.e., when the Constituents
have equal rapidity.   Thus the heaviest constituents have the highest momentum fractions
and highest $x$. Intrinsic charm thus predicts that the charm structure function has
support at large $x_{bj}$  in excess of DGLAP extrapolations~\cite{Brodsky:1980pb}; this
is in agreement with the EMC measurements~\cite{Harris:1995jx}.  It predicts leading
charm hadron production and fast charmonium production in agreement with
measurements~\cite{Anjos:2001jr}.   The production cross section for the double charmed
$\Xi_{cc}^+$ baryon~\cite{Ocherashvili:2004hi} and the production of double $J/\psi's$
appears to be consistent with the dissociation and coalescence of double IC Fock
states~\cite{Vogt:1995tf}.  Intrinsic charm can also explain the $J/\psi \to \rho \pi$
puzzle~\cite{Brodsky:1997fj}.  It also affects the extraction of suppressed CKM matrix
elements in $B$ decays~\cite{Brodsky:2001yt}.

\section{Solving for Light-front Wavefunctions}

In principle, one can solve for the LFWFs directly from the
fundamental theory using methods such as discretized light-front
quantization (DLCQ), the transverse lattice, lattice gauge theory
moments, or Bethe--Salpeter techniques.  DLCQ has been remarkably
successful in determining the entire spectrum and corresponding LFWFs in
1+1 field theories, including supersymmetric examples.  Reviews of
nonperturbative light-front methods may be found in
references~\cite{Brodsky:1997de,cdkm,Dalley:ug,Brodsky:2003gk}.  One can
also project the known solutions of the Bethe--Salpeter equation to equal
light-front time, thus producing hadronic light-front Fock wave
functions.  A potentially important method is to construct the $q\bar q$
Green's function using light-front Hamiltonian theory, with DLCQ boundary
conditions and Lippmann-Schwinger resummation.  The zeros of the
resulting resolvent projected on states of specific angular momentum
$J_z$ can then generate the meson spectrum and their light-front Fock
wavefunctions.  The DLCQ properties and boundary conditions allow a
truncation of the Fock space while retaining the kinematic boost and
Lorentz invariance of light-front quantization.

Even without explicit solutions, much is known about the explicit
form and structure of LFWFs.  They can  be matched to
nonrelativistic Schrodinger wavefunctions at soft scales.  At high
momenta, the LFWFs large $k_\perp$ and $x_i \to 1$ are constrained
by arguments based on conformal symmetry, the operator product
expansion, or perturbative QCD.  The pattern of higher Fock states
with extra gluons is given by ladder relations.  The structure of
Fock states with nonzero orbital angular momentum is also
constrained.

\section{The Infrared Behavior of  Effective QCD Couplings}

Theoretical~\cite{vonSmekal:1997is,Zwanziger:2003cf,%
Howe:2002rb,Howe:2003mp,Furui:2003mz} and
phenomenological~\cite{Mattingly:ej,Brodsky:2002nb,Baldicchi:2002qm}
evidence is now accumulating that the QCD coupling becomes
constant at small virtuality; {\em i.e.}, $\alpha_s(Q^2)$ develops
an infrared fixed point in contradiction to the usual assumption
of singular growth in the infrared.  If QCD running couplings are
bounded, the integration over the running coupling is finite and
renormalon resummations are  not required.  If the QCD coupling
becomes scale-invariant in the infrared, then elements of
conformal theory~\cite{Braun:2003rp} become relevant even at
relatively small momentum transfers.

Menke, Merino, and Rathsman~\cite{Brodsky:2002nb} and I have
presented a definition of a physical coupling for QCD which has a
direct relation to high precision measurements of the hadronic
decay channels of the $\tau^- \to \nu_\tau {\rm h}^-$.  Let
$R_{\tau}$ be the ratio of the hadronic decay rate to the leptonic
one.  Then $R_{\tau}\equiv
R_{\tau}^0\left[1+\frac{\alpha_\tau}{\pi}\right]$, where
$R_{\tau}^0$ is the zeroth order QCD prediction, defines the
effective charge $\alpha_\tau$.  The data for $\tau$ decays is
well-understood channel by channel, thus allowing the calculation
of the hadronic decay rate and the effective charge as a function
of the $\tau$ mass below the physical mass.  The vector and
axial-vector decay modes which can be studied separately.  Using an
analysis of the $\tau$ data from the OPAL
collaboration~\cite{Ackerstaff:1998yj}, we have found that the
experimental value of the coupling $\alpha_{\tau}(s)=0.621 \pm
0.008$ at $s = m^2_\tau$ corresponds to a value of
$\alpha_{\MSbar}(M^2_Z) = (0.117$-$0.122) \pm 0.002$, where the
range corresponds to three different perturbative methods used in
analyzing the data.  This result is in good agreement with the
world average $\alpha_{\MSbar}(M^2_Z) = 0.117 \pm 0.002$.  However,
one also finds that the effective charge only reaches
$\alpha_{\tau}(s) \sim 0.9 \pm 0.1$ at $s=1\,{\rm GeV}^2$, and it
even stays within the same range down to $s\sim0.5\,{\rm GeV}^2$.
The effective coupling is close to constant at low scales,
suggesting that physical QCD couplings become constant or
``frozen" at low scales.

The near constancy of the effective QCD coupling at small scales
helps explain the empirical success of dimensional counting rules
for the power law fall-off of form factors and fixed angle
scaling.  As shown in the
references~\cite{Brodsky:1997dh,Melic:2001wb}, one can calculate
the hard scattering amplitude $T_H$ for such
processes~\cite{Lepage:1980fj} without scale ambiguity in terms of
the effective charge $\alpha_\tau$ or $\alpha_R$ using
commensurate scale relations.  The effective coupling is evaluated
in the regime where the coupling is approximately constant, in
contrast to the rapidly varying behavior from powers of
$\alpha_{\rm s}$ predicted by perturbation theory (the universal
two-loop coupling).  For example, the nucleon form factors are
proportional at leading order to two powers of $\alpha_{\rm s}$
evaluated at low scales in addition to two powers of $1/q^2$; The
pion photoproduction amplitude at fixed angles is proportional at
leading order to three powers of the QCD coupling.  The essential
variation from leading-twist counting-rule behavior then only
arises from the anomalous dimensions of the hadron distribution
amplitudes.

Parisi~\cite{Parisi:zy} has shown that perturbative QCD becomes a
conformal theory  for $\beta \to 0$ and zero quark mass.  There are
a number of useful phenomenological consequences of near conformal
behavior: the conformal approximation with zero $\beta$ function
can be used as template for QCD
analyses~\cite{Brodsky:1985ve,Brodsky:1984xk} such as the form of
the expansion polynomials for distribution
amplitudes~\cite{Braun:2003rp,Braun:1999te}.  The near-conformal
behavior of QCD is also the basis for commensurate scale
relations~\cite{Brodsky:1994eh} which relate observables to each
other without renormalization scale or scheme
ambiguities~\cite{Brodsky:2000cr}.  An important example is the
generalized Crewther relation~\cite{Brodsky:1995tb}.  In this
method the effective charges of observables are related to each
other in conformal gauge theory; the effects of the nonzero QCD
$\beta-$ function are then taken into account using the BLM
method~\cite{Brodsky:1982gc} to set the scales of the respective
couplings.     The magnitude
of the corresponding effective charge~\cite{Brodsky:1997dh}
$\alpha^{\rm exclusive}_s(Q^2) = {F_\pi(Q^2)/ 4\pi Q^2 F^2_{\gamma
\pi^0}(Q^2)}$ for exclusive amplitudes is connected to
$\alpha_\tau$ by a commensurate scale relation.  Its magnitude:
$\alpha^{\rm exclusive}_s(Q^2) \sim 0.8$ at small $Q^2,$  is
sufficiently large as to explain the observed magnitude of
exclusive amplitudes such as the pion form factor using the
asymptotic distribution amplitude.

\section{AdS/CFT and Near-Conformal Field Theory}

As shown by Maldacena~\cite{Maldacena:1997re}, there is a remarkable correspondence
between large $N_C$ supergravity theory in a higher dimensional  anti-de Sitter space and
supersymmetric QCD in 4-dimensional space-time.  String/gauge duality provides a
framework for predicting QCD phenomena based on the conformal properties of the ADS/CFT
correspondence.  For example, Polchinski and Strassler~\cite{Polchinski:2001tt} have
shown that the power-law fall-off of hard exclusive hadron-hadron scattering amplitudes
at large momentum transfer can be derived without the use of perturbation theory by using
the scaling properties of the hadronic interpolating fields in the large-$r$ region of
AdS space.  Thus one can use the Maldacena correspondence to compute the leading
power-law falloff of exclusive processes such as high-energy fixed-angle scattering of
gluonium-gluonium scattering in supersymmetric QCD.   The resulting predictions for
hadron physics effectively coincide~\cite{Polchinski:2001tt,Brower:2002er,Andreev:2002aw}
with QCD dimensional counting
rules~\cite{Brodsky:1973kr,Matveev:ra,Brodsky:1974vy,Brodsky:2002st}.) Polchinski and
Strassler~\cite{Polchinski:2001tt} have also derived counting rules for deep inelastic
structure functions at $x \to 1$ in agreement with perturbative QCD
predictions~\cite{Brodsky:1994kg} as well as Bloom-Gilman exclusive-inclusive duality. An
interesting point is that the hard scattering amplitudes which are normally or order
$\alpha_s^p$ in PQCD appear as order $\alpha_s^{p/2}$ in the supergravity predictions.
This can be understood as an all-orders resummation of the effective
potential~\cite{Maldacena:1997re,Rey:1998ik}. The near-conformal scaling properties of
light-front wavefunctions thus lead to a number of important predictions for QCD which
are normally discussed in the context of perturbation theory.

De Teramond and I~\cite{Brodsky:2003px} have shown how one can use the scaling properties
of the hadronic interpolating operator in the extended AdS/CFT space-time theory to
determine the scaling of light-front hadronic wavefunctions at high relative transverse
momentum. The angular momentum dependence of the light-front wavefunctions also follow
from the conformal properties of the AdS/CFT correspondence.  The scaling and conformal
properties of the correspondence leads to a hard component of the light-front Fock state
wavefunctions of the form:
\[
\psi_{n/h} (x_i, \vec k_{\perp i} , \lambda_i, l_{z i}) \sim
\frac{(g_s~N_C)^{\frac{1}{2} (n-1)}}{\sqrt {N_C}}\\
 ~\prod_{i =1}^{n
- 1} (k_{i \perp}^\pm)^{\vert l_{z i}\vert}\\
~ \left[\frac{ \Lambda_o}{ {M}^2 - \sum _i\frac{\vec k_{\perp i}^2
+ m_i^2}{x_i} + \Lambda_o^2}  \right] ^{n +\vert l_z \vert -1}\ ,
\label{eq:lfwfR}
\]
where $g_s$ is the string scale and $\Lambda_o$ represents the
basic QCD mass scale.  The scaling predictions agree with the
perturbative QCD analysis given in the references~\cite{Ji:bw},
but  the AdS/CFT analysis is performed at strong coupling without
the use of perturbation theory.  The form of these
near-conformal wavefunctions can
be used as an initial ansatz for a variational treatment of the
light-front QCD Hamiltonian.

\section*{Acknowledgements}

I wish to thank Ayse Kizilersu, Tony Thomas, Tony  Williams and their colleagues at the
CSSM in Adelaide for  hosting this outstanding meeting. This talk is based on
collaborations with Guy de Teramond, John Hiller, Dae Sung Hwang, Volodya Karmanov, Sven
Menke, Carlos Merino, Jorg Raufeisen, and Johan Rathsman.  This work was supported by the
U.S. Department of Energy, contract DE--AC03--76SF00515.

\end {document}